\documentstyle[11pt,aaspp4]{article}

\slugcomment{To appear in the Dec 2000 issue of the Astronomical Journal}

\lefthead{}
\righthead{}
\def\deg{\mbox{$^\circ$}}

\newcommand{\kms}{$\,{\rm km\,s^{\scriptscriptstyle -1}}$}
\newcommand{\gtsim}{\ {\raise-0.5ex\hbox{$\buildrel>\over\sim$}}\ }
\newcommand{\ltsim}{\ {\raise-0.5ex\hbox{$\buildrel<\over\sim$}}\ }

\begin{document}

\title{Dynamical Constraints on the Formation of NGC~4472 and
Its Globular Clusters}

\author{Stephen E. Zepf}
\affil{Department of Astronomy, Yale University, New Haven, CT 06520; \\
zepf@astro.yale.edu}
\author{Michael A. Beasley\altaffilmark{1}}
\affil{Department of Physics, University of Durham, South Road, Durham DH1 3LE,
England; m.a.beasley@durham.ac.uk}
\author{Terry J. Bridges\altaffilmark{1}}
\affil{Anglo-Australian Observatory, Epping, NSW, 1710, Australia; 
tjb@aaoepp.aao.gov.au}
\author{David A. Hanes\altaffilmark{1}}
\affil{Department of Physics, Queen's University, Kingston, ON K7L 3N6, Canada;
hanes@astro.queensu.ca}
\author{Ray M. Sharples\altaffilmark{1}}
\affil{Department of Physics, University of Durham, South Road, Durham DH1 3LE,
England; r.m.sharples@durham.ac.uk}
\author{Keith M. Ashman}
\affil{Department of Physics and Astronomy, University of Kansas, Lawrence,
 KS 66045; ashman@kuspy.phsx.ukans.edu}
\author{Doug Geisler}
\affil{Grupo de Astronomia, Universidad de Concepci\'on, Casilla 160-C,
Concepci\'on, Chile; doug@kukita.cfm.udec.cl}

\altaffiltext{1}{Visiting Astronomers, Canada-France-Hawaii Telescope, 
which is operated by the National Research Council of Canada, 
the Centre National de la Recherche Scientifique, and the
University of Hawaii.} 

\begin{abstract}
	We present new radial velocities for 87 globular clusters
around the elliptical galaxy NGC~4472, and combine these with our 
previously published data to create a data set of velocities for 
144 globular clusters around NGC~4472. We utilize this data set 
to analyze the kinematics of the NGC~4472 globular cluster system.
The new data confirms our previous discovery that the metal-poor 
clusters have significantly higher velocity dispersion than the 
metal-rich clusters in NGC~4472.
We also find very little or no rotation in the more
spatially concentrated metal-rich population, 
with a resulting upper limit for this population
of $(v/\sigma)_{proj} < 0.34$ at a $99\%$ confidence level. 
The very small angular momentum
in the metal-rich population requires efficient angular momentum
transport during the formation of this population which is
spatially concentrated and chemically enriched.
Such angular momentum transfer can be provided by galaxy mergers,
but has not been achieved in other extant models of elliptical 
galaxy formation that include dark matter halos.
We also calculate the velocity dispersion as a function of
radius, and show that it is consistent with roughly isotropic
orbits for the clusters and the mass distribution of NGC~4472
inferred from X-ray observations of the hot gas around the galaxy.

\end{abstract}


\section{Introduction}

	Although early-type galaxies make up much of the stellar mass 
in the local universe (e.g.\ Fukugita, Peebles, \& Hogan 1998, 
Coppi \& Zepf 2000), there is no consensus on when or how they 
formed. The possibilities range from monolithic collapse at
high redshift to formation via galaxy mergers over a range
of redshifts. Many of the integrated properties of elliptical 
galaxies can be accounted for by both scenarios, making it
difficult to determine the formation histories of these galaxies.
For example, one of the salient observational features of
the population of early-type galaxies is that their color-magnitude 
and Mg$_2 - \sigma$ relations have small scatter. However,
both early collapse models (e.g.\ Kodama \& Arimoto 1997) and
hierarchical merging models (e.g.\ Kauffmann \& Charlot 1998)
have been shown to be consistent with the observed correlations.
Both mergers and collapse models can also be made consistent 
with current observations of galaxies at high redshift. 
The small number of very red galaxies in deep K-selected samples 
indicates that most ellipticals formed some of their stars at 
$z \ltsim 3$ (Treu \& Stiavelli 1999, Barger et al.\ 1999, Zepf 1997),
with a few exceptions (e.g.\ Treu et al.\ 1998, Benitez et al.\ 1999).
However, this only requires that some of the formation activity 
of ellipticals take place at lower redshifts, and does not
explicitly favor one model or another.

       A critical observation that must be addressed by any
successful model of galaxy formation is the weak or absent stellar 
rotation in giant ellipticals. In contrast to the dynamical 
insignificance of rotation in elliptical 
galaxies, the dynamics of spiral galaxies are dominated by 
rotation, resulting in about an order of magnitude more total 
angular momentum for spirals compared to the visible components
of ellipticals (e.g.\ Silk \& Wyse 1993). The large difference 
in angular momentum between elliptical and spiral galaxies has long
been recognized as one of the most fundamental problems in
galaxy formation. Moreover, this difference between the
angular momenta of spiral and elliptical galaxies
is not expected to be true of their dark matter halos.
All extant theoretical models predict that
the spin-up of galactic halos through tidal torques will be
similar for all galaxies, so the range of halo angular momenta is 
smaller than the observed difference between elliptical 
and spiral galaxies (e.g.\ Heavens \& Peacock 1988, Barnes \& 
Efstathiou 1987,  Warren et al.\ 1992, Eisenstein \& Loeb 1995).
In terms of the commonly used dimensionless spin parameter
$\lambda = LE^{1/2}/G M^{5/2}$, 
where $L$ is the angular momentum
E is the energy, and M is the mass, 
all halos have $\lambda \simeq 0.06$.
These calculations also find that the angular momentum 
of a halo is mostly independent of its mass or environment, 
so any differences between the angular momentum of elliptical 
and spiral galaxies can not be accounted for solely 
by their different masses or environments. 

	The rotation and density of spiral galaxies can be
straightforwardly understood as the result of the collapse and spin-up of 
baryonic material inside a dark matter halo (e.g.\ Fall \& Efstathiou 1980).
More specifically, for a protogalaxy with $\lambda \sim 0.06$,
a collapse factor of roughly ten produces both the observed rotation 
and the density of spiral disks. 
However, there is no such simple explanation for the low angular 
momentum in the visible parts of elliptical galaxies. 
As described above, calculations of the tidal torques on galaxy halos 
show that all halos have fairly similar specific angular momentum. 
Moreover, the similar stellar densities of ellipticals and spirals 
suggest that the gas in each galaxy type collapsed by roughly the 
same degree during the galaxy formation process. The comparable dark matter
contributions within the visible regions of each galaxy type, as well
as the extended nature of dark matter halos around both spirals and 
ellipticals, also imply that the gas collapse factors in both 
morphological types must have been similar. Since the collapse
factors seem to be similar, and the initial values of $\lambda$
are expected to be similar, the dramatic difference in the
rotation in the visible regions of ellipticals and spirals
is a critical constraint on the galaxy formation process.
Specifically, spirals and ellipticals must have had different formation 
processes, with efficient angular momentum transport occurring 
during the formation of elliptical galaxies.

	It has long been realized that galaxy mergers might provide
a mechanism for transporting angular momentum out of the visible
regions of elliptical galaxies (e.g.\ Fall 1979, Toomre 1977,
Barnes \& Hernquist 1992). 
The hypothesis that 
elliptical galaxies form from major mergers while spiral galaxies 
have avoided a major merger in the recent past is therefore an 
attractive one. However, as described above, 
it is difficult to demonstrate from observational data that 
elliptical galaxies today formed from mergers in the past.
Some of the strongest evidence that elliptical galaxies
form episodically comes from the study of the colors of globular 
clusters around nearby ellipticals. These studies show that elliptical 
galaxies often have globular cluster systems with bimodal metallicity 
distributions (Ashman \& Zepf 1998 and references therein, see also
Kundu 1999 and Gebhardt \& Kissler-Patig 1999). 
The distinct globular cluster populations observed in many
ellipticals suggest episodic formation and are inconsistent with 
simple monolithic collapse models.
In contrast, bimodal globular cluster systems were predicted
by Ashman \& Zepf (1992) for elliptical galaxies that formed from 
mergers of gas-rich spirals. Specifically, Ashman \& Zepf (1992) 
predicted that elliptical galaxies would have a population of metal-poor
globulars from the halos of the progenitor spirals, and a population
of metal-rich clusters formed in the merger that made the elliptical.
Subsequent observations have produced a large body of observational 
evidence indicating that globular clusters can form in gas-rich mergers
(see reviews by Schweizer 1998, Ashman \& Zepf 1998, and references therein).
Thus both fundamental predictions of the merger model have been
supported by subsequent observations - globular clusters are observed 
to form in mergers and the proposed product of these mergers,
elliptical galaxies, show evidence for multiple populations of
globular clusters.

	The discovery of young globular clusters in mergers and of 
bimodality in the globular cluster systems of ellipticals
were successfully predicted by the merger models. However,
for many elliptical galaxies the specific number and inferred 
metallicity of the blue, metal-poor globular clusters is not in 
detailed agreement with the predictions of the simple merger model.
In the simplest merger picture, these metal-poor clusters
come from the halo populations of the progenitor spirals,
and therefore the metal-poor population in ellipticals should 
have the same metallicity and specific frequency
(number per stellar luminosity, or preferably, number
per stellar mass).
However, some elliptical galaxies such as M87 appear to have 
more metal-poor clusters than can be accounted for by combining 
the halo populations of spirals like the Galaxy and M31
(e.g. Forbes, Brodie, \& Grillmair 1997, Lee, Kim, \& Geisler 1998) 
while other ellipticals such as NGC~3923 (Zepf, Ashman, \& Geisler 1995)
and IC~4051 (Woodworth \& Harris 2000)
may either not have a
substantial metal-poor globular cluster population or have 
one which is more metal-rich than the halo populations of 
the Galaxy and M31. Zepf et al.\ (1995)
and Ashman \& Zepf (1998) have noted that, in the 
context of the merger models, these observations suggest
variations in the properties of the globular cluster
systems of the progenitor spirals, as well as the possibility
of significant accretion of low-metallicity globular clusters 
and associated dwarf galaxies.

	Following the discovery of the bimodality in the
globular cluster systems of elliptical galaxies, models
were constructed to try to account for bimodality without
major mergers. One example is
the proposal that elliptical galaxies simply form in two phases, 
an initial spatially extended phase of metal-poor globular 
globular cluster formation followed by the later formation 
of metal-rich clusters after dissipation and enrichment, 
with some internal mechanism
that turns off globular cluster formation between the 
two globular cluster formation phases (Forbes et al.\ 1997,
Harris, Harris, \& McLaughlin 1998). 
Alternatively, it has also been proposed that the central
component of elliptical galaxies undergoes standard dissipative
collapse, and the metal-poor globular cluster population
is accreted later (Cote, Marzke, \& West 1999).

	The key physical element of all models is that angular 
momentum conservation requires the dissipative collapse that 
leads to the formation of the metal-rich globular clusters
to also result in significant rotation in the metal-rich 
globular cluster population. The only way to avoid this
conclusion is to have a mechanism to efficiently transport
angular momentum outwards. Merger models have been shown
to provide such a mechanism, while smooth dissipative
collapse models do not naturally transport angular
momentum efficiently. Therefore, the kinematics of the globular
cluster systems provides a way to test if the physics of
the formation of elliptical galaxies and their globular
cluster systems is consistent with smooth dissipative collapse
or if angular momentum is required to be transported outwards.
Moreover, if angular momentum transport is indicated by observations 
in the inner regions of an elliptical, it might be possible to observe 
the angular momentum transferred outwards in the
kinematics at very large radii, as possibly found for the
M87 globular cluster system by Kissler-Patig \& Gebhardt (1998) 
based on the data of Cohen \& Ryzhov (1997). 

	Measurements of the radial velocities of globular
clusters also provide information about the mass distribution
of the galaxy and the orbits of the clusters. Globular
clusters are particularly useful for studying the dynamics
of the outer halos of elliptical galaxies because they
can be observed out to much larger radii than it is possible 
to obtain spectroscopy of the integrated light.
A very large number of velocities are required to
independently determine the mass distribution and the orbits
of the tracer particles in a completely non-parametric way 
(e.g.\ Merritt \& Tremblay 1994). However, if the mass distribution 
inferred from X-ray observations and the assumption of hydrostatic 
equilibrium in the hot gas is adopted, the orbits of the 
globular clusters can be constrained.
Conversely, if assumptions are made about the cluster
orbits (e.g.\ that they are isotropic), then the mass distribution 
can be estimated. In practice, a sensible approach is to
check for consistency of the mass distribution determined
via the X-ray observations of the hot gas with dynamical
measurements given simplifying assumptions about the orbits,
as each technique has its own systematic concerns which are
mitigated if the independent approaches agree. 

	In this paper, we present the results of a spectroscopic
survey of globular clusters around the Virgo elliptical NGC~4472.
We focus on the analysis of the kinematics of the globular
cluster system and its implications. A discussion of the ages
and metallicities of the NGC~4472 globular clusters is given
in Beasley et al.\ (2000). The observations and data reduction 
are described in $\S 2$. The results of our analysis of these 
data are presented in $\S 3$. In $\S 3.1$ we compare the velocity 
dispersions of the different globular cluster populations, 
in $\S 3.2$ we analyze the rotation of the cluster populations, 
in $\S 3.3$ we study the radial trends in the velocity dispersion and
rotation, and in $\S 3.4$ we examine the mass distribution of NGC~4472
inferred from our data. The implications of these results are 
discussed in $\S 4$, and our conclusions given in $\S 5$.

\section{Observations and Data Reduction}

	We obtained spectra of globular cluster candidates around 
NGC~4472 using the Multi-Object Spectrograph (hereafter MOS,
Crampton et al.\ 1992, LeFevre et al.\ 1994) on the Canada-France- 
Hawaii Telescope over parts of three nights in May 1998.
MOS is an effective instrument for this study because its
relatively large field of view allows for efficient multiplexing
of the NGC~4472 globular cluster system, which extends out to 
many arcminutes from the center of the galaxy 
(cf. Rhode \& Zepf 2000).
The candidate globular clusters were selected from the photometric 
study of the NGC~4472 globular cluster system by Geisler et al.\ (1996).
We obtained spectra through four slitmasks, each of which contained
slitlets for approximately 45 globular cluster candidates.
We preferentially targeted cluster candidates with magnitudes
between $ 19.5 < V < 21.5$, but included fainter clusters when
useful for maximizing the number of targets in each mask.
About 2/3 of the slitlets yielded useful spectra, with the remainder 
having either low signal-to-noise or problematic sky subtraction.

	The spectra were obtained with a 600 l/mm grism, giving 
a resolution of 2.2 \AA\ per pixel, and a spectral range of 
$3800 - 6500 \AA\ $. The data were processed with flat fields
taken at the beginning and end of the night, and wavelength
calibrated using frequent Hg arcs. After extraction of the 
object spectra and sky subtraction, the radial velocities of 
the candidate globular clusters were determined by cross-correlation 
(e.g.\ Tonry \& Davis 1979) with six template stars of known velocity
and spectral types ranging from F8V to K0III. 
The `r' value of the cross-correlation was required to be
greater than 2.5 for a velocity measurement to be included
in our final sample. The final velocity for each object was
determined by a weighted average of the velocities returned 
by the cross-correlation with each of the six template stars,
where the weighting was given by the cross-correlation peak
height of each template. The data reduction 
procedures are described more completely in Beasley et al.\ (2000).

	The formal errors returned by the cross-correlation
task for each globular cluster are given in Table 1 below.
These are typically 50-100 \kms\ . A check on these errors
can be made by comparing the velocities for the 13 clusters
for which we have velocities from both this analysis and
our earlier work (Sharples et al.\ 1998, hereafter S98). 
The result is a mean difference of
$-17$ \kms\ with an uncertainty in any individual measurement
of 78 \kms\ . Much of the dispersion is driven by a large
offset for one cluster, so the typical uncertainty may
be smaller. Regardless, these uncertainties are much smaller
than the velocity dispersion of NGC~4472, so the measurement
errors in the velocities of individual clusters do not have 
a significant effect on the results.

\section{Results}

	In Table 1, we list the positions, magnitudes, colors, and 
radial velocities for the 144 clusters around NGC~4472 for which 
we have spectra. This table includes 100 new globular cluster 
velocities obtained in the observations described above, and 57 
previously published by us (S98), 
which also included some from earlier work by Mould et al.\ (1990).
For the 13 clusters with data from both datasets, the tabulated
velocity is from the observation with the strongest cross-correlation.
These data form the basis for the analysis presented in the remainder 
of this paper. For completeness, we also list in Table 2 
the 20 objects for which we obtained spectra and found 
them to be foreground stars or background galaxies. 
Our $83\%$ success rate in spectroscopically confirming
our targets as globular clusters is higher than earlier work.
This high success rate is likely due to selection of
targets from high quality photometry, and bodes well for
future spectroscopic programs.

	In addition to analyzing the populations as a whole, 
we also investigate the kinematic properties of the metal-rich 
and metal-poor populations of globular clusters in NGC~4472. 
These populations were first identified through the discovery
of bimodality in the color distribution of the NGC~4472 globular
cluster system (Zepf \& Ashman 1993 and many subsequent papers). 
That these color differences are primarily due to metallicity
is now confirmed by spectroscopy (Beasley et al.\ 2000).
Further evidence for metallicity driving the color distribution
is provided by analysis of the luminosity functions of the red 
and blue globular clusters (e.g.\ Puzia et al.\ 1999, Lee \& Kim 2000).
In order to divide our sample into populations of metal-rich 
and metal-poor clusters, we apply the KMM mixture-modeling
algorithim to the full dataset of Geisler et al.\ (1996). 
This objective analysis divides the sample at a color of
$(C-T_1) = 1.625$. Clusters bluer than this are more likely to belong
to the metal-poor population while those redder than this value are
more likely to belong to the metal-rich population. This is the same
analysis and color cut used in S98. This assignment represents 
the best statistical estimate of whether a globular cluster 
belongs to the metal-rich or metal-poor population.
However not every cluster will necessarily
be assigned to the population to which it truly belongs
because of possible overlap in the color distributions 
of the different cluster populations as well as photometric errors. 
This will tend to make it more difficult to detect physical differences 
between the metal-rich and metal-poor cluster populations,
particularly for clusters near the boundary.

	The full dataset is plotted in Figure 1. This 
figure shows the distribution on the sky of the globular 
clusters for which we have velocities, and also includes 
a rough indication of the velocities of these objects,
as well as their identification as a member of the metal-rich
or metal-poor cluster population. The bulk of this paper is 
aimed at a quantitative analysis of the velocities and positions 
of the clusters presented in Table 1 and Figure 1. 

\subsection{Velocity Dispersions}

	One of the goals of this study was to compare the kinematics 
of the metal-rich and metal-poor globular cluster populations
of NGC~4472. In our earlier work with radial velocities for 57 globulars 
we found tentative evidence ($86\%$ confidence level) for a higher
velocity dispersion in the metal-poor population compared to
the metal-rich population (S98).
A difference in velocity dispersions between the metal-rich and metal-poor
globular cluster populations is critical evidence that these populations 
have real physical differences, since the division between the 
metal-rich and metal-poor cluster populations was made by an objective 
analysis of the colors alone.

	Figure 2 shows the comparison of the velocity distribution
of the red, metal-rich clusters, and the blue, metal-poor clusters
around NGC~4472. This plot demonstrates that the new, larger dataset 
strongly confirms our original discovery that the metal-poor cluster 
population has a significantly higher velocity dispersion than the 
metal-rich cluster population. Specifically, a F-test indicates that 
the two populations have a different dispersion at $> 99.9\% $ 
confidence level. The values of the velocity dispersion in each population 
are $\sigma =  221 \pm 22$ \kms\ for the metal-rich clusters, and 
$\sigma = 356 \pm 25$ \kms\ for the metal-poor clusters 
where the uncertainties have been determined from bootstrapping.
These are uncorrected for any rotation, but as we show below,
the rotation is small and does not affect the dispersions significantly.
Moreover, regardless of its origin, the difference in velocity dispersions 
provides dynamical evidence for the original distinction between 
the metal-rich and metal-poor cluster populations in NGC~4472.

	It is also interesting to note that the difference in dispersions 
between the metal-rich and metal-poor globular cluster populations
in NGC~4472 does not necessarily imply that the populations have different 
orbital properties. This is because there is observational evidence that 
the metal-rich cluster system in NGC~4472 is more spatially concentrated 
than the metal-poor cluster system (e.g.\ Lee et al.\ 1998). 
This difference between the density profiles of the metal-poor and 
metal-rich globular cluster populations makes it possible for these 
populations to have similar orbital anisotropy but different velocity 
dispersions, and still follow the same gravitational potential.
In detail, the velocity dispersion difference we observe is
modestly larger ($\sim 2\sigma$) than that expected from the 
difference in the spatial distributions for the metal-poor
and metal-rich clusters found by Lee et al.\ (1998).

\subsection{Rotation}
	
	A second major goal of our spectroscopic study of the NGC~4472 
globular cluster system was to determine the rotation of the system 
as a whole, as well as that of the metal-rich and metal-poor populations.
As discussed in the introduction, determining the rotation of
the cluster populations in NGC~4472 is invaluable for addressing
questions about the formation history of the galaxy and the relationship
of the different cluster populations to one another.

	A straightforward way to estimate the rotation in the globular 
cluster system of NGC~4472 is to do a non-linear least-squares fit to 
the equation
\begin{equation}
V(r)=V_{rot}\sin (\theta-\theta_0)+V_o .
\end{equation}
This determines the best fitting flat-rotation
curve with the position angle of the line of nodes ($\theta_0$)
and the rotation velocity ($V_{rot}$) free parameters.
The results from this analysis are given in Table 3.
We find little or no rotation in the metal-rich population and
modest evidence for rotation in the full sample and
in the metal-poor population. The line of nodes of the best fitting
rotation solution appears to be similar to the position
angle of the major axis of the galaxy ($PA \simeq 172\deg$).
The significance of the rotation detections
can be tested by performing Monte Carlo simulations of the data.
These simulations retain the position angles of the objects but 
randomize the velocities (cf.\ S98).  
Roughly half of such simulations give a rotation
signature about some axis equal to that observed in the metal-poor
and total cluster samples. However, the rotation in these
simulations is not preferentially aligned with the major
axis of the galaxy, as the real data is. 
We can account for this by considering random 
samples with the position angle of the line of nodes fixed to be 
that observed for the isophotes the galaxy.
In this case, only $16\%$ of the random simulations give a 
rotation signature as large as observed for the total and metal-poor
cluster samples. The metal-rich sample still does not show evidence
for rotation even when the position angle is fixed.

	In addition to the best fitting rotation solution, we
determine an upper limit on the rotation by creating
Monte Carlo samples of clusters with the same number
of objects and the same velocity dispersion as the given
populations, but with a given rotation velocity. We define the
$99\%$ confidence upper limit as the rotation velocity for
which only 1/100 of such random samples gives a rotation
signature as small as that observed. This procedure produces
strict upper limits on the rotation in the globular cluster
system of NGC~4472, especially the metal-rich population.
Specifically, the $99\%$ upper limit on rotation for 
the full cluster sample is 120 \kms\ , for the metal-poor cluster
population the $99\%$ upper limit on rotation is 200 \kms, 
and for the metal-rich population the
rotation is less than 75 \kms\ at $99\%$ confidence.



	These limits on rotation can be used to set upper limits
on the ratio of $(v/\sigma)_{proj}$, which is a common diagnostic of
the importance of rotation. For the metal-rich population, the combination
of the $99\%$ upper limit of 75 \kms\ on the rotation with the
velocity dispersion of 221 \kms\ found earlier gives an upper
limit of $(v/\sigma)_{proj} < 0.34$ at 99\% confidence.
This is remarkably low compared to the expectation of simple
collapse models. It is also dramatically lower than the metal-rich
thick disk/bulge clusters of the Milky Way, which have $v/\sigma > 1$.
Although the value for NGC~4472 is a projected one, it is
very unlikely that rotation is dynamically significant.
One way to see this is to compare the observed $(v/\sigma)_{proj}$
of the metal-rich cluster system in NGC~4472 to that expected for 
an oblate rotator (e.g.\ Binney \& Tremaine 1987). For the observed 
ellipticity of about 0.2 for the metal-rich population
(Lee et al.\ 1998, Rhode \& Zepf 2000),
an oblate rotator would have a $(v/\sigma)_{proj}$ of about 0.4,
mostly independent of projection. This prediction for an oblate
rotator is larger than the $99\%$ upper limit, strongly suggesting 
that the metal-rich globular cluster system of NGC~4472 is not supported 
by rotation. A similar approach is to note that the inclination would 
have to be $i < 12\deg$ for $(v/\sigma)_{proj} = 0.34$ to be
consistent with $v/\sigma \ge 1$. Moreover, such an inclination 
would require an intrinsic flattening of greater than E9 for the 
metal-rich system to match the observed E2 in projection.

	The very small rotation in the metal-rich globular
cluster system of NGC~4472 is problematic for models in which 
central spheroidal systems all form similarly, with the only difference 
being the mass of the central bulge/elliptical component. Instead,
it requires efficient angular momentum transport for the metal-rich
globular cluster population of NGC~4472 that did not occur
during the formation of the metal-rich globular cluster
population of the Milky Way. Such a difference is natural 
in models in which elliptical galaxies like NGC~4472 formed
in major mergers, and spiral galaxies formed in such a way
as to avoid disruptive events that lead to effective, large scale 
transport of angular momentum. This is discussed in more detail
in the final section of the paper.

\subsection{Radial Variations of Rotation and Velocity Dispersion}

	In addition to the total velocity dispersions and rotation 
velocities derived in the two previous sections, radial trends in
these quantities are also interesting. For example, if the velocity 
dispersion can be traced out to large radii, constraints can be placed
on the mass distribution of the halo, although some simplifying 
assumptions about the orbital anisotropy of the globular cluster 
population are required for modest sample sizes. Conversely, a mass 
distribution can be adopted based on other data such as X-ray observations, 
and then the observed dispersions can be used to constrain the orbital
anisotropy at different radii. Similarly, changes in the rotation
velocity as a function of radius can be used to trace the distribution
of angular momentum within a galaxy.

	In order to determine the rotational velocity and velocity
dispersion as a function of radius, it is necessary to average
in some way the discrete radial velocities of individual clusters.
Binning is not ideal because of its well-known sensitivity to
the choice of bin centers and sizes. The problem of estimating
smooth functions from individual datapoints is a well-developed
statistical field (e.g. Simonoff 1996 and references therein). 
The application of these techniques to the problem of estimating 
the velocities and dispersion of the globular cluster systems as 
a function of radius is straightforward because our goal is to 
determine the overall trends in these systems and sharp changes 
are not expected. In this case, a Gaussian smoothing kernel provides 
reliable results with the advantage of simplicity in application and 
interpretation.

	A key question for any smoothing procedure is to determine
the width of the smoothing kernel. A common approach, called
cross-validation, is to determine the kernel size that minimizes 
the summed total of the variances determined at each point, when 
that point is excluded from the calculation. However, cross-validation
gives little guidance when the sample in question is consistent
with being a constant, because it is statistically preferable
in this case to use all of the data to determine the value of
the function in question at each radius. This is exactly the
case for the velocities of the globular cluster system of
NGC~4472, particularly the full sample and the metal-poor
subsample. Therefore, while cross-validation clearly identifies
the statistical consistency of the data with a flat rotation
curve, it does not yield much help for determining the appropriate
smoothing kernel for determining any dependence of the
velocity dispersion on radius. Because our goal is to determine
the overall trend of velocity dispersion with radius,
we adopt a fairly broad smoothing, using a Gaussian kernel
with $\sigma = 100''$. This kernel width is roughly
equivalent to two or three independent radial ``bins.''
Smaller smoothing kernels produce the same overall trends,
with additional bumps and wiggles. A sign that these may
be undersmoothed is that the velocity dispersion profiles
from smaller kernels yield unphysical results when used
in the Jeans equation (see Section 3.4). 

	Figure 3 shows the results of our analysis of
the radial variation of the rotation and velocity dispersion.
Because the line of nodes of rotation is consistent with the 
position angle of the galaxy (Section 3.2), the data have been 
flipped about the minor axis to take full advantage of the 
available positions and velocities.
The upper and lower lines are $1\sigma$ bootstrap limits.
In order to build the bootstrap samples in a way that is
consistent with the smoothing kernel used to determine the
best fit, the probability of including an object in the bootstrap 
sample is proportional to the value of the kernel of that
object at a given radius.
Because of the smoothing required to get rotation and velocity
dispersions from individual radial velocities, the
individual points along the curves are of course not independent
of one another.

	Overall, the globular cluster population in NGC~4472
appears to have a slowly declining velocity dispersion with radius.
This is seen in the full sample, and the individual metal-rich
and metal-poor samples are consistent with the same trend. 
A sharper drop in the dispersion at our largest radial distances
($r \gtsim 400''$) is also consistent with the data, as shown
by the lower limit of the dispersion in Figure 3 and also by
analyses using smaller smoothing kernels. Further data at large
radii is required to test this possibility.
The slightly declining velocity dispersion in NGC~4472 contrasts
to the slightly rising dispersion found for M87 over the
same radial range (e.g.\ Cohen 2000 and references therein). 
This difference may suggest that the rising dispersion in M87 
is due to transition from the galaxy to the Virgo cluster itself,
as M87 is located at the center of the Virgo cluster,
while NGC~4472 is well away from the cluster center.

	Figure 3 also shows the rotational velocity of the
total sample and the metal-poor population is consistent with 
a constant value at all radii, while the metal-rich sample shows 
slight evidence for increasing rotation outwards. This
can also be seen by performing a straight regression of
velocity with major axis distance, which gives absolutely
no trend for the total and metal-poor samples, and a $1\sigma$
result for increasing rotation with larger radii for
the metal-rich sample. At all radii for which we have data the 
rotation is significantly smaller than the dispersion. This 
applies to the globular cluster sample as a whole, as well as 
to the metal-rich and metal-poor cluster populations.

\subsection{M/L ratio and Orbital Isotropy}

	The velocity dispersion profile determined in
the previous section can be used to derive the mass distribution
of NGC~4472 through the application of the Jeans equation
(e.g. Binney \& Tremaine 1987).
The mass determined in this way depends on the anisotropy
of the cluster orbits, as well as on the directly observed
projected velocity dispersion and projected density profiles.
The simplest approach is to adopt isotropy for the cluster
orbits, and to see how this compares to other estimates for
the mass of NGC~4472 within the same radial range.
In detail, the three dimensional luminosity density and velocity
dispersion profiles are determined from the observed projected 
quantities by numerically integrating the Abel equations.
Spherical symmetry is assumed in this approach, which is
not strictly true for NGC~4472, but it not likely to be
too far wrong given that the observed ellipticity of the 
globular cluster systems is $0.1-0.2$ (e.g.\ Lee et al.\ 1998).

	Figure~4 shows the mass of NGC~4472 as a function of radius.
The lines are masses inferred from application of the Jeans
equation to the radial velocities of our globular cluster sample. 
The central solid line is the best fit to the 144 radial velocities 
discussed in this paper. The dotted lines are the $1\sigma$ lower 
and upper limits. These uncertainties in the mass are based 
on the bootstrapped uncertainties in the radial velocity
dispersion profile given in the previous section. We do not
include uncertainties in the density profile of the globular
cluster system because they are much smaller than those for the
velocity dispersion. Specifically, the surface density profile
is taken from the fit of a de Vaucouleurs' law to the wide field
CCD imaging study of Rhode \& Zepf (2000), which extends out to 
$\gtsim 20'$, well beyond our last data point. 
Different techniques for deriving the density profile
from these data or comparisons to previous work
produce variations that are much smaller than the uncertainty 
in the velocity dispersion.

	The radial profile of the mass distribution of NGC~4472
derived from the radial velocities of the globular clusters can be
compared to the mass inferred from X-ray observations
of the hot gas around the galaxy. The solid squares on Figure~4
are the masses inferred from ROSAT observations of the 
hot gas around NGC~4472 (Irwin \& Sarazin 1996).
The open squares represent points for which the assumption
of hydrostatic equilibrium on which the X-ray masses are
based may be uncertain because the X-ray isophotes are
irregular at these radii. 

	The overall agreement between the mass inferred from the 
globular cluster velocities given isotropic orbits and the 
mass inferred from the analysis of X-ray observations of 
the hot gas assuming hydrostatic equilibrium is good.
This agreement suggests that both assumptions are probably roughly 
correct, and that the masses so derived are reasonably accurate.
The conclusion that then follows is that the mass-to-light
ratio of NGC~4472 is at least five times greater at
radii of $\sim 30$ kpc than it is at radii of several kpc,
indicative of a substantial dark halo around this galaxy.
This is some of the strongest dynamical evidence to date
for massive halos around elliptical galaxies.
At small radii, there is a hint that the mass inferred from 
the globular cluster dynamics may be slightly higher than that 
inferred from the hot gas. This could either be the result of 
statistical uncertainties in either observation, a slight radial 
anisotropy in the globular cluster orbits at smaller radii, 
or a slight underestimation of the temperatures in the hot gas 
at small radii. In any case the overall agreement between
the two approaches is good, and more data will be required
to extend this approach to larger radii and to test for
any differences at smaller radii.

\section{Implications}

	One of the primary goals of the study of the kinematics
of the globular cluster systems of nearby ellipticals is to 
constrain the formation history of these galaxies. The bimodal
color distributions of the globular cluster systems of
many ellipticals already indicate an episodic formation history
for these galaxies, roughly along the lines predicted by merger 
models. The kinematic data presented here confirm and extend
the identification of two populations of globular clusters 
in NGC~4472 by showing that the population of globular clusters 
identified as metal-poor through their colors has a higher 
velocity dispersion than the population identified as
metal-rich. 

	In addition to confirming the existence of a metal-rich
and metal-poor globular cluster populations in NGC~4472, our
data help constrain the physical nature of the episodic formation
history of this galaxy. Perhaps the most critical observational
result is that there is little or no rotation in the metal-rich 
globular cluster system. Despite the absence of rotation,
the metal-rich system clearly underwent significant dissipation 
and collapse since it is more spatially concentrated than the 
metal-poor cluster population and much more so than the dark 
matter halo. Given conservation of angular momentum, the
metal-rich population would be naturally expected to spin
up as it collapses. In fact, significant rotation is observed
observed in the metal-rich population of disk galaxies like
the Milky Way and M31. Thus the minimal rotation in the
metal-rich globular cluster population of NGC~4472 requires 
angular momentum transport during the formation of this
elliptical galaxy, and distinguishes the metal-rich 
globular cluster population in this elliptical from
metal-rich cluster populations in the Galaxy and M31.
Merger models have long been put forward as a way to
transport angular momentum outwards (e.g.\ Toomre \& Toomre 1972,
Barnes \& Hernquist 1992 and references therein).
A generally consistent picture can be constructed in which
elliptical galaxies like NGC~4472 form in major mergers which
create the metal-rich globular cluster population and transfer 
angular momentum outwards, while disk galaxies like the Milky Way 
and M31 have only had more minor mergers, which may lead to modest 
amounts of globular cluster formation but which are not as efficient 
at angular momentum transfer.

	An additional implication of our data is that NGC~4472
has a dark halo that extends out to the limit of our dataset,
about $30$ kpc. However, this limit is only set by the field
of view of the spectrographs we have used. A wide-field CCD 
imaging study of the globular cluster system of NGC~4472 shows 
that it extends out to at least $80$ kpc (Rhode \& Zepf 2000). 
Thus globular clusters offer a dynamical probe of the halo at very 
large radii which can be reached effectively with spectrographs 
with sufficient field of view. This is particularly valuable 
for NGC~4472, since its X-ray isophotes are irregular at radii 
greater than about $20$ kpc, indicating the hot gas may not be 
in hydrostatic equilibrium. Without hydrostatic equilibrium,
mass estimates from X-ray gas are problematic, leaving dynamical
techniques the best hope for extending the understanding of the
halo of NGC~4472 to very large radii.

\section{Conclusions}

	In this paper, we present radial velocities for 144 globular 
clusters around NGC~4472. Based on this dataset, we analyze the
kinematics of the globular cluster system as a function radius 
and metallicity. We find

1. The metal-poor globular clusters have a significantly higher
velocity dispersion than the metal-rich clusters. This kinematic
difference confirms the identification of these two cluster populations 
within NGC~4472, as they were originally selected based on their 
colors alone.

2. $(v/\sigma)_{proj}$ is much less
than one over the radial range covered by our data ($\sim 3-30$ kpc). 
This is true for the globular cluster system as a whole and for the 
individual metal-rich and metal-poor cluster systems. The upper
limit on rotation is particularly small for the metal-rich globular
cluster system, for which $(v/\sigma)_{proj} < 0.34$ at the 
$99\%$ confidence level.

3. This absence of rotation requires either significant
angular momentum transport outwards during the formation 
of NGC~4472 or a much lower initial spin of the halo
of NGC~4472 than given by standard tidal torque calculations.
The requirement for significant angular momentum transport
is particularly strong for the metal-rich population,
which is concentrated towards the center of the galaxy
and has a high metallicity. These properties point to significant 
dissipation and collapse during the formation of the metal-rich 
cluster population, yet it has little or no rotation.

4. The absence of significant rotation in the metal-rich globular
cluster population of NGC~4472 strongly distinguishes this metal-rich
cluster population from the metal-rich cluster populations
found in disk galaxies like the Milky Way. Thus the angular
momentum transport required for the metal-rich cluster
population in NGC~4472 appears to be associated with the physical 
differences between elliptical and spiral galaxies. This may be 
explained if ellipticals form from major mergers, while spirals 
are those galaxies that have not had major mergers since the 
formation of their disk.

5. The velocity dispersion profile of the NGC~4472 slowly
declines with radius, with a hint that the decline may steepen
at large radius. The mass derived from this dispersion profile
and the assumption of isotropic orbits is consistent with
that derived from X-ray observations of the hot gas. This
suggests that neither technique has dramatic systematic
problems, and strongly indicates the presence of a dark matter
halo around NGC~4472.

\acknowledgements

	We thank Dave Carter for his contributions to the
early stages of the research project, David Buote for
making available his analysis of the ROSAT observations of 
NGC~4472, and the referee for a careful reading of the paper.
We acknowledge useful conversations with Ken Freeman, Melinda Weil, 
and Dean McLaughlin.
SEZ acknowledges support from the Hellman Family Foundation,
MAB acknowledges the support of a PPARC studentship and the
use of STARLINK facilities at the University of Durham,
DG acknowledges support for this project from CONICYT
through Fondecyt grant 1000319 and by the Universidad de Concepci\'on
through research grant No.\ 99.011.025-1.0, and
and DAH acknowledges support from an Operating Grant
awarded by the Natural Sciences and Engineering Research
Council (NSERC) of Canada.

\clearpage

\begin{figure}
\plotone{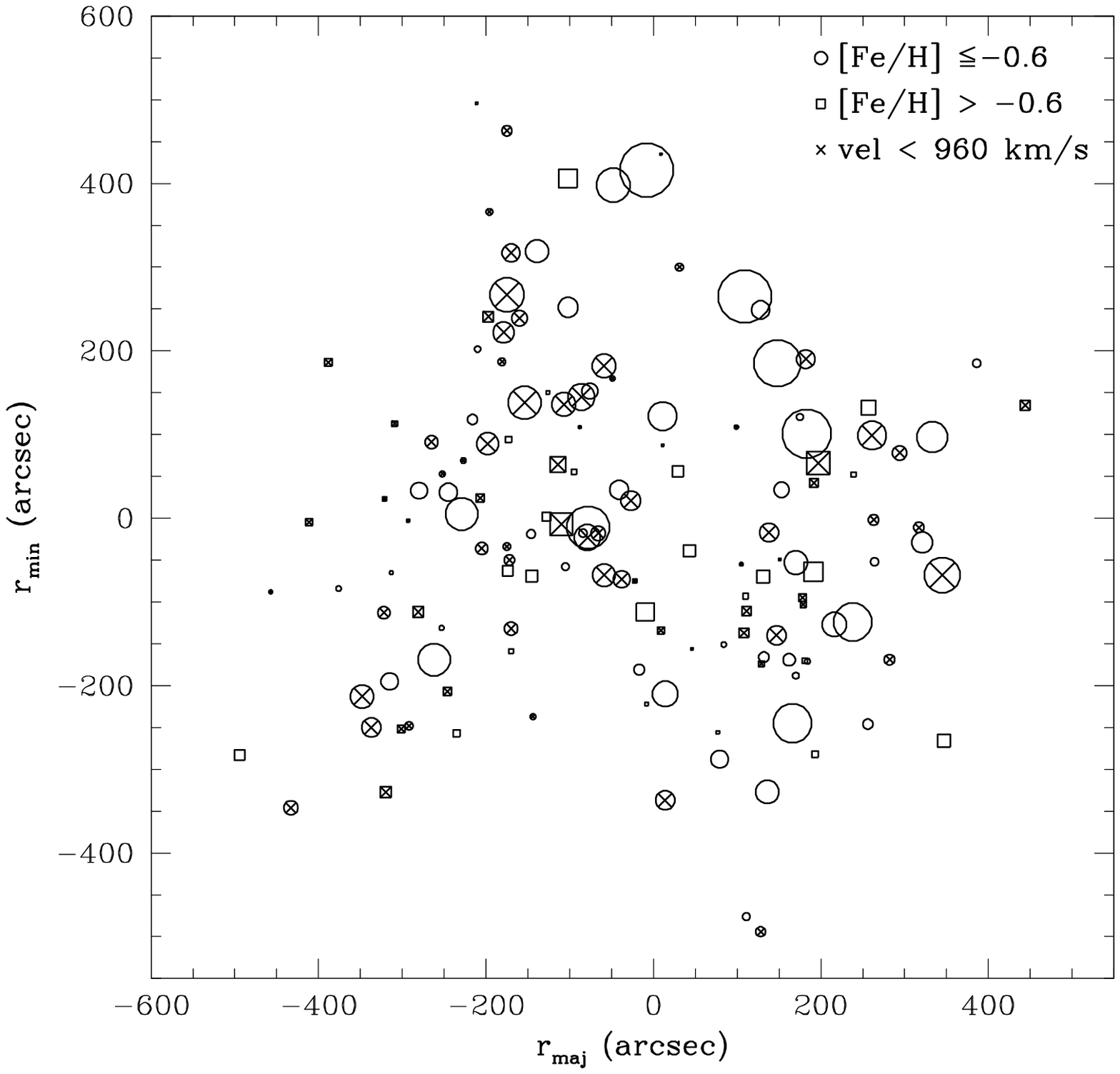}
\vskip -0.6in
\caption{A plot of the location and velocities for our
NGC~4472 globular cluster dataset. The clusters are plotted
at their location on the sky, with a symbol size proportional to
their velocity difference from the systemic velocity of NGC~4472. 
An $x$ indicates a negative velocity with respect to the velocity 
of the galaxy, and an open symbol indicates a positive velocity.
Globular clusters identified by their colors as part of the metal-rich
population are plotted as squares, and globular clusters in the
metal-poor population are plotted as circles. This plot shows that
the dispersion falls off gradually with radius, and that there is
no obvious rotation about any axis, especially for the metal-rich
cluster population.}
\end{figure}

\begin{figure}
\plotone{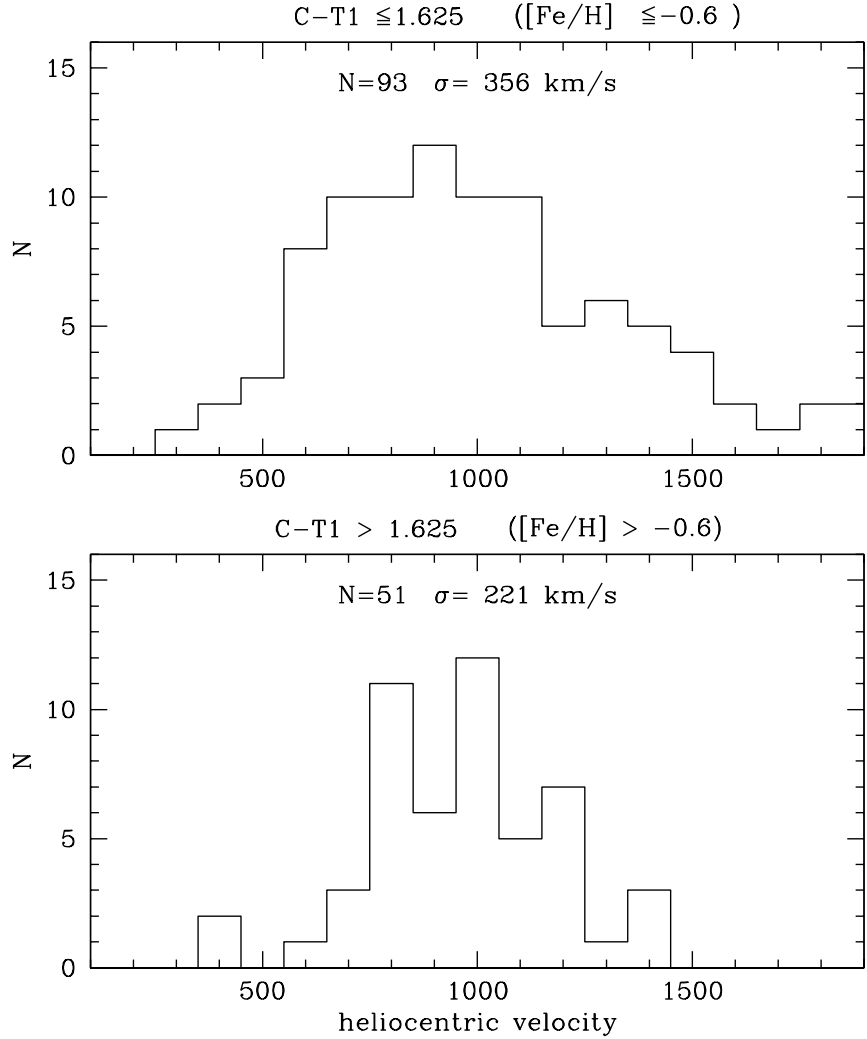}
\vskip -0.2in
\caption{Histograms of the velocities for the metal-poor (upper panel)
and metal-rich (lower panel) globular cluster populations in NGC~4472.
The larger velocity dispersion of the metal-poor clusters is clearly
evident. A F-test indicates that this difference is significant at
greater than $99\%$. This provides independent confirmation of the
original identification of metal-rich and metal-poor populations
of globular clusters in this galaxy.}
\end{figure}

\begin{figure}
\plotone{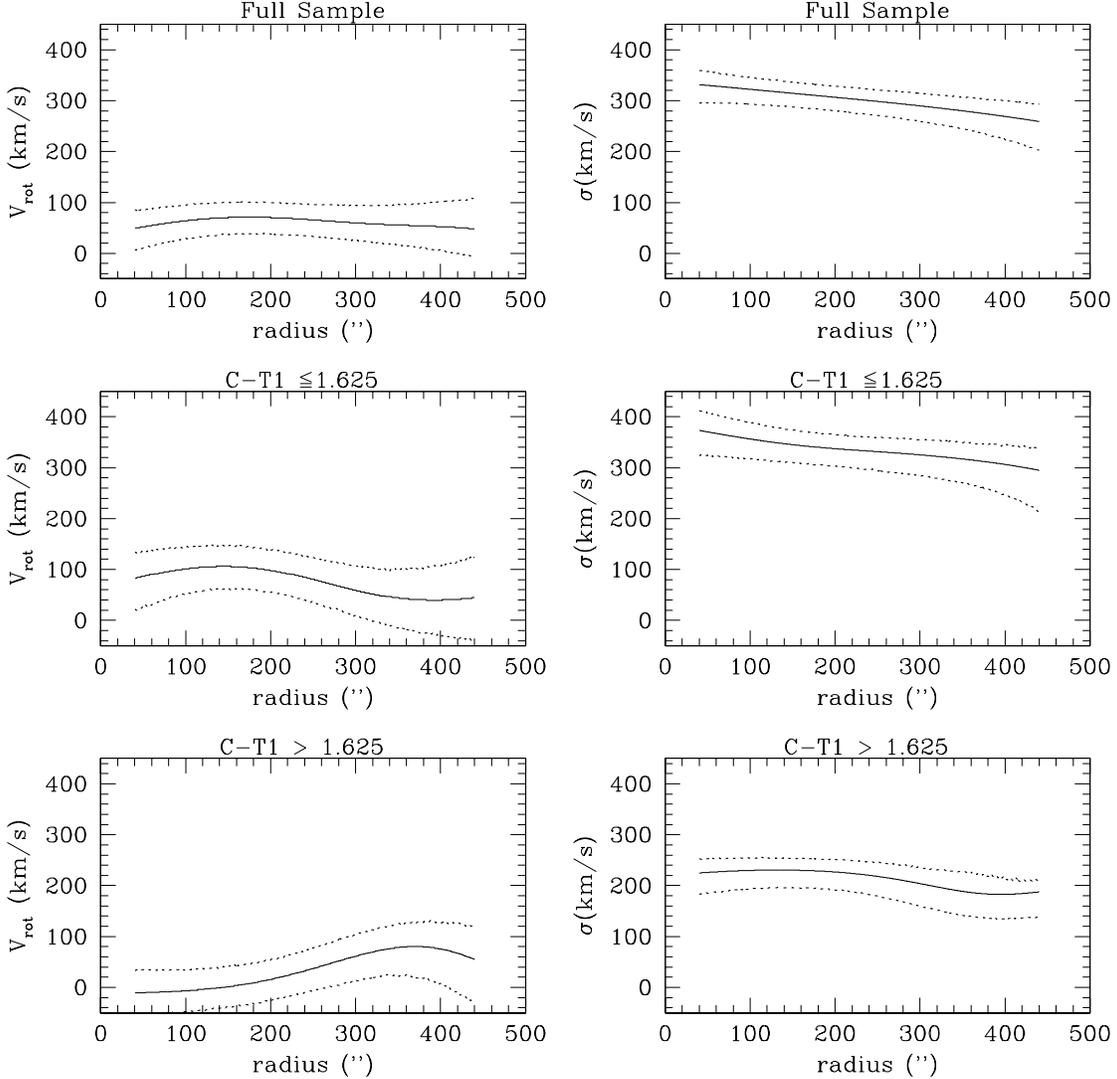}
\vskip -0.2in
\caption{Plots of the rotation and velocity dispersion
fields for the globular clusters of NGC~4472. 
The top panels are for the full data set, the middle
panels for the metal-poor (blue) clusters and the bottom panels
for the metal-rich (red) clusters. A Gaussian kernel with
$\sigma = 100''$ was used for the radial smoothing for all of
the datasets. The dotted lines show the $1\sigma$ uncertainties,
as determined from bootstrapping. The curves are highly correlated
in the radial direction with the smoothing used.
The plots show modest rotation in the full sample and
the metal-poor cluster population which is essentially constant 
with radius. The red sample has essentially zero rotation
at small radius and a tentative ($1\sigma$) rise to modest
rotation at larger radii. The velocity dispersion
is signficantly larger than the rotation at all radii.}
\end{figure}

\begin{figure}
\plotone{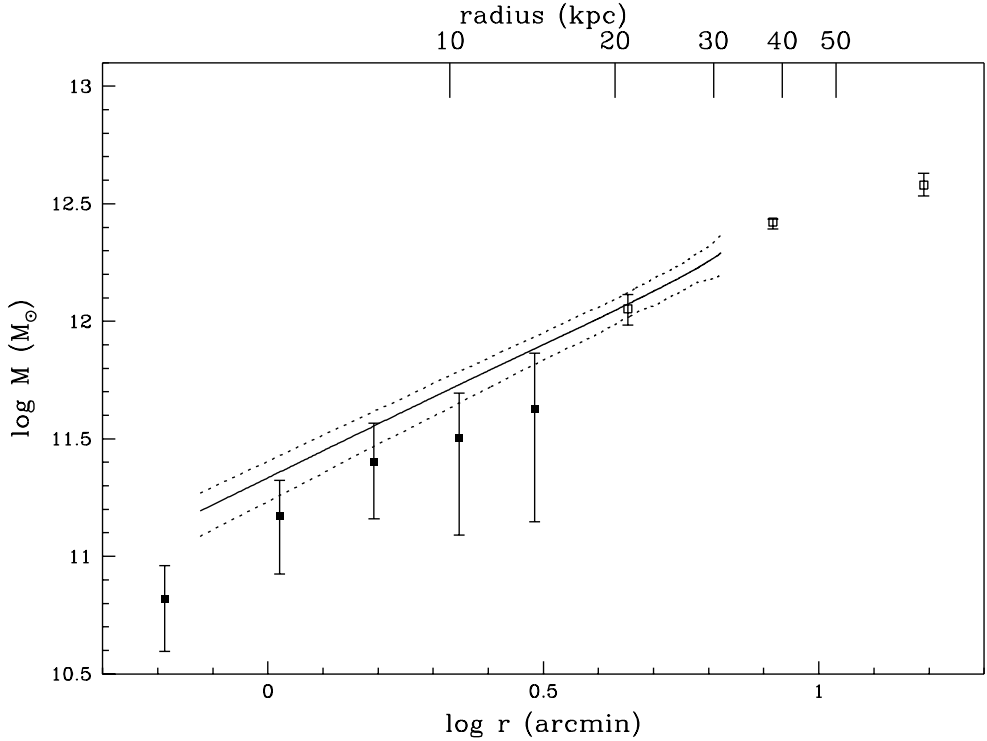}
\vskip -0.6in
\caption{A plot of the mass of NGC~4472 as a function of radius.
The lines are masses inferred from the radial velocities
of the globular clusters. The central solid line is the best
fit to the 144 radial velocities discussed in this paper.
The dotted lines are the $1\sigma$ lower and upper limits
determined via bootstrapping. 
All of these are based on the assumption of isotropic orbits
for the globular clusters.
The points are masses inferred from ROSAT observations of the 
hot gas around NGC~4472 (Irwin \& Sarazin 1996).
The open squares represent points for which the assumption
of hydrostatic equilibrium on which the X-ray masses are
based may be uncertain because the X-ray isophotes are
irregular at these radii. The overall agreement between
the masses inferred from the two techniques is good,
suggesting that the assumptions underlying each approach
are probably roughly correct. The conclusion that then follows 
is that NGC~4472 has a substantial dark halo, with a mass-to-light 
ratio at several tens of kpc that is at least a factor of five 
greater than in the inner regions of the galaxy.}
\end{figure}

\clearpage

\begin{deluxetable}{cccccc}

\tablecaption{Velocities of Globular Clusters in NGC~4472}
\tablehead{
\colhead {ID\tablenotemark{1}} & \colhead{R.A.} & \colhead{Decl.} &
\colhead{T$_1$} & \colhead{C$-$T$_1$} & V$_{hel}$ \\ 
\colhead{} & \colhead{(B1950.0)} & \colhead{(B1950.0)} &
\colhead{(mag)} & \colhead{(mag)} & \colhead{(\kms)}
}
\startdata
 170 & 12 27 05.19 & 8 09 52.6 & 21.19 & 1.73 & 1243 $\pm  69$ \nl
 282 & 12 27 32.88 & 8 10 24.1 & 20.38 & 1.65 &  751 $\pm  39$ \nl
 463 & 12 27 17.87 & 8 10 55.5 & 19.93 & 1.59 &  342 $\pm  27$ \nl
 637 & 12 27 10.03 & 8 11 24.1 & 19.95 & 1.54 &  814 $\pm  31$ \nl
 647 & 12 27 04.56 & 8 11 25.4 & 21.15 & 1.34 & 1101 $\pm  39$ \nl
 676 & 12 27 19.80 & 8 11 30.5 & 20.72 & 1.46 & 1304 $\pm  56$ \nl
 714 & 12 27 34.88 & 8 11 35.4 & 20.32 & 1.55 & 1061 $\pm  39$ \nl
 744 & 12 27 20.91 & 8 11 40.2 & 19.73 & 1.38 &  814 $\pm  29$ \nl
 876 & 12 27 28.14 & 8 11 58.3 & 19.63 & 1.46 & 1487 $\pm  24$ \nl
 995 & 12 26 46.05 & 8 12 10.0 & 20.90 & 1.54 &  828 $\pm  90$ \nl
1047 & 12 27 00.94 & 8 12 13.8 & 21.14 & 1.89 & 1083 $\pm  30$ \nl
1087 & 12 27 17.20 & 8 12 17.5 & 19.64 & 1.44 & 1070 $\pm  29$ \nl
1110 & 12 27 12.04 & 8 12 20.3 & 19.88 & 1.37 & 1626 $\pm  24$ \nl
1207 & 12 27 26.09 & 8 12 29.3 & 21.34 & 1.25 &  739 $\pm  69$ \nl
1234 & 12 26 46.82 & 8 12 32.2 & 20.81 & 1.47 & 1059 $\pm  64$ \nl
1255 & 12 27 20.37 & 8 12 34.4 & 19.71 & 1.35 &  816 $\pm  66$ \nl
1315 & 12 27 11.39 & 8 12 40.9 & 20.68 & 1.42 & 1370 $\pm  76$ \nl
1423 & 12 27 02.74 & 8 12 51.3 & 20.94 & 1.57 & 1641 $\pm  29$ \nl  
1448 & 12 26 56.90 & 8 12 54.2 & 20.79 & 1.34 & 1353 $\pm  26$ \nl  
1475 & 12 27 07.89 & 8 12 57.0 & 21.15 & 1.46 & 1018 $\pm  37$ \nl  
1518 & 12 27 07.89 & 8 13 00.6 & 19.25 & 1.85 & 1050 $\pm  36$ \nl  
1570 & 12 27 06.52 & 8 13 05.6 & 20.98 & 1.58 & 1034 $\pm  61$ \nl  
1587 & 12 27 26.80 & 8 13 07.3 & 21.16 & 1.12 &  471 $\pm  75$ \nl  
1650 & 12 27 23.30 & 8 13 13.6 & 20.85 & 1.95 & 1040 $\pm  55$ \nl  
1712 & 12 27 07.54 & 8 13 19.0 & 20.36 & 1.34 & 1144 $\pm  40$ \nl  
1731 & 12 27 28.81 & 8 13 21.1 & 20.71 & 1.82 & 1294 $\pm  51$ \nl  
1749 & 12 27 14.86 & 8 13 22.6 & 20.92 & 1.98 & 1407 $\pm  88$ \nl
1764 & 12 27 12.15 & 8 13 23.6 & 20.82 & 1.72 &  855 $\pm  37$ \nl
1798 & 12 27 12.65 & 8 13 25.9 & 20.69 & 1.98 &  811 $\pm  31$ \nl
1982 & 12 27  9.12 & 8 13 42.0 & 20.89 & 1.01 &  648 $\pm  43$ \nl
2031 & 12 27 15.13 & 8 13 46.4 & 20.71 & 1.37 & 1352 $\pm  65$ \nl
2045 & 12 27  6.50 & 8 13 47.8 & 20.94 & 1.77 &  857 $\pm  54$ \nl
2060 & 12 27  7.15 & 8 13 48.3 & 20.62 & 1.29 & 1108 $\pm  65$ \nl
2140 & 12 27 21.65 & 8 13 55.5 & 20.45 & 1.80 &  784 $\pm  31$ \nl
2163 & 12 27 23.34 & 8 13 57.7 & 20.15 & 2.01 &  402 $\pm  43$ \nl
2195 & 12 26 58.16 & 8 14 00.8 & 21.33 & 1.14 & 1241 $\pm  56$ \nl
2256 & 12 27 14.99 & 8 14 05.8 & 21.24 & 1.90 &  954 $\pm  28$ \nl
2341 & 12 27 00.22 & 8 14 13.0 & 20.76 & 1.91 & 1001 $\pm  68$ \nl
2406 & 12 27 13.24 & 8 14 18.5 & 20.84 & 2.03 & 1244 $\pm  70$ \nl
2420 & 12 27 08.45 & 8 14 19.3 & 20.95 & 2.08 &  763 $\pm  92$ \nl
2452 & 12 27 25.28 & 8 14 22.5 & 21.49 & 1.40 & 1828 $\pm  81$ \nl
2482 & 12 27 10.16 & 8 14 24.7 & 21.58 & 2.08 &  767 $\pm  56$ \nl
2528 & 12 27 16.79 & 8 14 28.3 & 20.34 & 1.46 &  654 $\pm  65$ \nl
2543 & 12 27 20.34 & 8 14 29.3 & 20.27 & 1.36 & 1199 $\pm  48$ \nl
2569 & 12 27 11.33 & 8 14 31.7 & 20.12 & 1.89 & 1056 $\pm  46$ \nl
2634 & 12 27 07.09 & 8 14 38.0 & 19.70 & 1.56 & 1014 $\pm  57$ \nl
2753 & 12 27 13.65 & 8 14 47.0 & 20.88 & 1.19 &  945 $\pm 100$ \nl
2759 & 12 26 53.73 & 8 14 47.3 & 19.97 & 1.31 &  654 $\pm  92$ \nl
2817 & 12 27 30.93 & 8 14 50.6 & 21.05 & 1.50 &  665 $\pm  47$ \nl
3150 & 12 27 05.95 & 8 15 12.2 & 21.40 & 1.79 &  952 $\pm  42$ \nl
3307 & 12 27 29.92 & 8 15 21.4 & 20.25 & 1.53 & 1790 $\pm  65$ \nl
3361 & 12 27 01.91 & 8 15 26.6 & 20.34 & 1.55 & 1392 $\pm  33$ \nl
3628 & 12 27 00.59 & 8 15 44.0 & 21.22 & 1.90 & 1008 $\pm  49$ \nl
3635 & 12 27 24.04 & 8 15 44.5 & 20.86 & 1.34 &  936 $\pm  94$ \nl
3757 & 12 27 13.40 & 8 15 51.8 & 21.02 & 1.82 & 1220 $\pm  95$ \nl
3808 & 12 27 06.60 & 8 15 54.6 & 20.35 & 1.83 &  832 $\pm  35$ \nl
3909 & 12 27 33.58 & 8 16 00.4 & 21.43 & 1.36 & 1253 $\pm  90$ \nl
3980 & 12 27 03.05 & 8 16 05.4 & 21.15 & 1.28 & 1112 $\pm  45$ \nl
4168 & 12 27 07.64 & 8 16 19.4 & 20.36 & 1.68 & 1384 $\pm  44$ \nl
4210 & 12 27 34.25 & 8 16 22.9 & 20.53 & 1.62 & 1910 $\pm  29$ \nl
4386 & 12 27 19.17 & 8 16 34.7 & 19.83 & 1.94 & 1197 $\pm  33$ \nl
4513 & 12 27 09.78 & 8 16 42.8 & 20.10 & 1.85 &  908 $\pm  80$ \nl
4731 & 12 27 09.56 & 8 16 58.2 & 19.96 & 1.43 &  698 $\pm  57$ \nl
4780 & 12 27 20.80 & 8 17 00.9 & 19.52 & 1.95 &  971 $\pm  45$ \nl
4959 & 12 27 23.06 & 8 17 11.9 & 21.38 & 1.33 & 1449 $\pm  44$ \nl
5090 & 12 27 09.43 & 8 17 20.3 & 19.83 & 1.61 &  582 $\pm  46$ \nl
5323 & 12 27 16.29 & 8 17 33.9 & 20.33 & 0.82 & 1263 $\pm  65$ \nl
5456 & 12 27 12.49 & 8 17 41.4 & 19.26 & 1.39 &  737 $\pm  65$ \nl
5561 & 12 26 56.85 & 8 17 48.3 & 20.82 & 1.39 &  903 $\pm  48$ \nl
5564 & 12 27 34.86 & 8 17 48.5 & 21.26 & 1.37 &  862 $\pm  19$ \nl
5629 & 12 27 11.90 & 8 17 52.3 & 21.09 & 1.36 &  522 $\pm  52$ \nl
5707 & 12 27 12.63 & 8 17 55.9 & 21.47 & 1.44 & 1712 $\pm  30$ \nl
5750 & 12 27 12.08 & 8 17 58.9 & 21.50 & 1.47 & 1063 $\pm  89$ \nl
5856 & 12 27 09.09 & 8 18 06.4 & 21.03 & 1.13 & 1050 $\pm  84$ \nl
6108 & 12 27 24.66 & 8 18 22.8 & 21.49 & 1.42 &  913 $\pm  39$ \nl
6164 & 12 27 12.26 & 8 18 27.2 & 19.79 & 1.65 &  426 $\pm  30$ \nl
6231 & 12 27 16.55 & 8 18 32.3 & 20.77 & 1.82 & 1046 $\pm  50$ \nl
6284 & 12 27 25.39 & 8 18 36.8 & 19.44 & 1.57 &  569 $\pm  54$ \nl
6294 & 12 27 01.30 & 8 18 37.7 & 21.02 & 1.64 & 1034 $\pm  84$ \nl
6344 & 12 27 07.59 & 8 18 41.3 & 20.89 & 2.01 & 1220 $\pm  50$ \nl
6357 & 12 27 20.14 & 8 18 41.9 & 20.47 & 1.26 &  958 $\pm  29$ \nl
6388 & 12 27 23.15 & 8 18 44.5 & 20.18 & 1.36 & 1212 $\pm  24$ \nl
6394 & 12 27 03.03 & 8 18 45.0 & 21.40 & 1.42 &  760 $\pm  85$ \nl
6427 & 12 27 12.47 & 8 18 47.5 & 21.11 & 1.79 & 1141 $\pm  50$ \nl
6476 & 12 27 43.03 & 8 18 50.2 & 20.94 & 2.23 &  966 $\pm  36$ \nl
6485 & 12 27 22.50 & 8 18 51.2 & 21.07 & 1.53 &  510 $\pm  82$ \nl
6520 & 12 27 16.76 & 8 18 53.6 & 20.06 & 1.86 &  607 $\pm  57$ \nl
6564 & 12 27 10.77 & 8 18 57.3 & 20.03 & 1.34 & 1077 $\pm  31$ \nl
6615 & 12 27 41.49 & 8 19 01.3 & 20.48 & 1.47 & 1923 $\pm  49$ \nl
6696 & 12 27 21.46 & 8 19 08.5 & 20.08 & 1.59 &  561 $\pm  23$ \nl
6701 & 12 26 53.65 & 8 19 08.6 & 20.97 & 2.00 & 1092 $\pm 141$ \nl
6721 & 12 27 07.36 & 8 19 10.5 & 21.09 & 1.73 & 1180 $\pm  45$ \nl
6748 & 12 27 08.24 & 8 19 12.4 & 20.21 & 1.53 &  817 $\pm  20$ \nl
6872 & 12 27 09.21 & 8 19 20.8 & 20.15 & 1.46 &  870 $\pm  41$ \nl
6989 & 12 27 22.00 & 8 19 30.6 & 20.61 & 1.75 & 1009 $\pm  24$ \nl
7028 & 12 27 39.52 & 8 19 33.5 & 21.40 & 1.38 & 1548 $\pm  39$ \nl
7043 & 12 26 56.69 & 8 19 34.4 & 20.47 & 1.78 &  808 $\pm  67$ \nl
7095 & 12 27 29.05 & 8 19 39.8 & 21.43 & 1.56 & 1285 $\pm  80$ \nl
7197 & 12 27 08.44 & 8 19 48.2 & 20.94 & 1.50 &  782 $\pm  50$ \nl
7281 & 12 27 20.64 & 8 19 53.8 & 21.43 & 1.25 &  389 $\pm 123$ \nl
7340 & 12 27 17.40 & 8 19 59.0 & 20.91 & 1.77 & 1067 $\pm  29$ \nl
7364 & 12 26 58.73 & 8 20 01.4 & 21.36 & 1.36 & 1522 $\pm 100$ \nl
7399 & 12 27 01.33 & 8 20 05.1 & 20.35 & 1.40 & 1005 $\pm  44$ \nl
7430 & 12 26 53.05 & 8 20 06.5 & 20.76 & 1.41 &  862 $\pm  66$ \nl
7449 & 12 26 47.44 & 8 20 07.7 & 20.75 & 1.65 &  724 $\pm  65$ \nl
7458 & 12 27 12.24 & 8 20 08.7 & 20.75 & 1.84 &  807 $\pm  57$ \nl
7531 & 12 26 52.55 & 8 20 13.5 & 19.71 & 2.07 &  818 $\pm  72$ \nl
7616 & 12 27 16.55 & 8 20 20.2 & 21.35 & 1.49 &  600 $\pm  90$ \nl
7659 & 12 27 10.56 & 8 20 24.0 & 19.87 & 1.34 & 1520 $\pm  44$ \nl
7702 & 12 27 38.87 & 8 20 27.4 & 21.19 & 1.88 & 1388 $\pm  58$ \nl
7746 & 12 27 27.01 & 8 20 31.5 & 21.29 & 1.42 &  712 $\pm  68$ \nl
7784 & 12 27 23.23 & 8 20 34.4 & 19.20 & 1.52 &  868 $\pm  51$ \nl
7798 & 12 27 32.56 & 8 20 35.4 & 20.96 & 1.39 & 1340 $\pm  39$ \nl
7872 & 12 27 14.71 & 8 20 42.0 & 20.33 & 1.45 &  908 $\pm  77$ \nl
7886 & 12 26 55.98 & 8 20 43.8 & 20.62 & 1.58 & 1236 $\pm  33$ \nl
7889 & 12 27 25.49 & 8 20 43.5 & 18.85 & 1.58 &  614 $\pm  65$ \nl
7894 & 12 27 01.99 & 8 20 37.3 & 21.61 & 1.73 &  730 $\pm  81$ \nl
7914 & 12 27 18.07 & 8 20 46.4 & 21.20 & 1.33 & 1101 $\pm  29$ \nl
7938 & 12 27 11.91 & 8 20 47.7 & 20.92 & 1.44 & 1251 $\pm  50$ \nl
7945 & 12 26 51.93 & 8 20 48.2 & 19.79 & 1.59 &  651 $\pm  33$ \nl
8000 & 12 27 28.43 & 8 20 53.3 & 21.08 & 1.49 &  368 $\pm  40$ \nl
8090 & 12 27 13.14 & 8 21 00.9 & 20.51 & 1.46 &  903 $\pm  66$ \nl
8143 & 12 27 31.74 & 8 21 04.7 & 20.96 & 1.52 &  672 $\pm 109$ \nl
8164 & 12 27 26.33 & 8 21 06.9 & 21.32 & 2.04 &  738 $\pm  40$ \nl
8165 & 12 27 23.58 & 8 21 06.9 & 20.22 & 1.39 & 1027 $\pm  47$ \nl
8210 & 12 26 54.15 & 8 21 10.1 & 20.43 & 1.32 &  576 $\pm  88$ \nl
8273 & 12 27 01.10 & 8 21 15.9 & 20.60 & 1.43 &  784 $\pm  90$ \nl
8332 & 12 27 11.29 & 8 21 21.1 & 21.09 & 1.40 & 1226 $\pm 100$ \nl
8353 & 12 27 08.71 & 8 21 22.8 & 20.03 & 1.98 &  928 $\pm  40$ \nl
8357 & 12 27 04.34 & 8 21 22.9 & 20.26 & 1.45 &  981 $\pm  61$ \nl
8384 & 12 27 15.30 & 8 21 24.4 & 21.39 & 1.41 &  768 $\pm  54$ \nl
8596 & 12 27 34.37 & 8 21 44.6 & 19.73 & 1.31 &  888 $\pm  31$ \nl
8653 & 12 26 43.83 & 8 21 49.5 & 20.50 & 1.26 &  744 $\pm  44$ \nl
8712 & 12 27 40.99 & 8 21 54.7 & 20.93 & 1.49 &  817 $\pm  63$ \nl
8740 & 12 27 09.77 & 8 21 57.5 & 21.27 & 2.14 &  913 $\pm 106$ \nl
8890  &12 27 15.78 & 8 22 14.1 & 20.41 & 1.88 &  870 $\pm  65$ \nl
8919 & 12 27 01.83 & 8 22 16.0 & 19.87 & 1.45 & 1014 $\pm  65$ \nl
9145 & 12 27 42.37 & 8 22 39.1 & 19.79 & 1.76 &  973 $\pm  38$ \nl
9360 & 12 26 46.61 & 8 23 06.8 & 21.00 & 1.67 & 1191 $\pm  99$ \nl
9414 & 12 27 06.09 & 8 23 14.3 & 20.82 & 1.74 &  832 $\pm  40$ \nl
9527 & 12 26 59.87 & 8 23 32.0 & 20.91 & 1.58 &  941 $\pm  61$ \nl
9666 & 12 27 18.87 & 8 23 51.3 & 20.04 & 1.74 &  811 $\pm  71$ \nl
9991 & 12 27 26.39 & 8 14 35.5 & 19.41 & 1.27 & 1040 $\pm  65$ \nl
9992 & 12 27 15.80 & 8 17 16.7 & 19.99 & 1.48 &  641 $\pm  65$ \nl
\enddata

\tablenotetext{1}{ID from Geisler et al.\ (1996).}
\end{deluxetable}

\clearpage

\begin{deluxetable}{cccccc}

\tablecaption{Foreground Stars and Background Galaxies near NGC~4472
Identified in MOS Spectra}
\tablehead{
\colhead {ID\tablenotemark{1}} & \colhead{R.A.} & \colhead{Decl.} &
\colhead{T$_1$} & \colhead{C$-$T$_1$} & \colhead{V$_{hel}$} \\ 
\colhead{} & \colhead{(B1950.0)} & \colhead{(B1950.0)} &
\colhead{(mag)} & \colhead{(mag)} & \colhead{(\kms)} 
}
\startdata
4415  & 12 27 09.11 & 08 16 36.7 & 20.45 & 2.17 & 45000 \nl 
2597  & 12 27 12.91 & 08 14 34.2 & 20.50 & 1.24 & 20000 \nl 
2431  & 12 27 19.22 & 08 14 20.4 & 21.44 & 1.28 & 14800 \nl 
1746  & 12 26 47.59 & 08 13 22.5 & 20.70 & 1.61 & 222   \nl 
2430  & 12 27 03.46 & 08 14 20.4 & 21.45 & 1.75 & $-60$ \nl 
7790  & 12 26 47.97 & 08 20 34.7 & 19.74 & 1.41 & 11334 \nl 
8113  & 12 27 06.93 & 08 21 02.1 & 20.88 & 1.33 & 10982 \nl 
7569  & 12 27 35.79 & 08 20 16.3 & 21.25 & 1.44 & 10706 \nl 
7039  & 12 27 43.90 & 08 19 34.2 & 21.21 & 1.70 & 34686 \nl 
8062  & 12 27 30.61 & 08 20 58.8 & 19.97 & 1.18 &   83  \nl 
9103  & 12 27 36.56 & 08 22 35.6 & 20.97 & 1.35 &  224  \nl 
3900  & 12 27 16.98 & 08 16 00.0 & 21.04 & 1.89 & 10600 \nl 
3217  & 12 27 36.55 & 08 15 16.3 & 21.05 & 1.29 & 8868  \nl 
1922  & 12 27 37.08 & 08 13 36.7 & 21.16 & 1.60 & 32278 \nl 
1024  & 12 27 14.18 & 08 12 12.5 & 20.99 & 1.11 &  200  \nl 
1621  & 12 27 27.58 & 08 13 11.0 & 19.69 & 1.41 &  159  \nl 
2935  & 12 27 40.19 & 08 14 57.6 & 19.87 & 1.06 &  133 \nl 
1955  & 12 27 11.63 & 08 13 39.4 & 21.04 & 1.20 & 31000 \nl 
8224  & 12 27 17.37 & 08 21 11.4 & 21.34 & 1.79 & 11700 \nl 
6302  & 12 27 33.74 & 08 18 38.2 & 20.49 & 1.95 & 12000 \nl 

\enddata

\tablenotetext{1}{ID from Geisler et al.\ (1996)}.
\end{deluxetable}

\clearpage 

\begin{table}
\begin{tabular}{cccccccc} 
\hline\hline
 & Sample & & $V_0$ & $V_{rot}$ & $\theta_0$ & $V_{rot}^{max}$ (99\%) & \\
 & & & (\kms) & (\kms) &(degrees) & (\kms) & \\
\hline
 & All clusters (N=144) & & 1018 & 69  & 181 & $<130$ & \\
 & Blue Clusters (N=93) & & 1059 & 101 & 177 & $<200$ & \\
 & Red Clusters (N=51)  & & 940  & 15  & 180 & $<75$  & \\
\hline
\end{tabular}
\caption{Results from non-linear fits of equation 1 to globular cluster 
samples taken from Table 1. The second column gives the mean velocity
for each sample. The third column gives the least-squares
fit to the rotation velocity and the fourth column gives the
best-fit position angle. The final column gives the $99\%$
upper limit on the rotational velocity.}
\end{table}

\end{document}